\documentclass[12pt]{article}
\usepackage{amsfonts}
\usepackage{amssymb}
\usepackage{amsmath}
\usepackage[T2A]{fontenc}
\usepackage[cp1251]{inputenc}
\usepackage{pstricks}
\usepackage{graphicx}
\usepackage{multicol}
\numberwithin{equation}{section}
\renewcommand{\Re}{\text{Re}}

\begin{document}

\title{Quantum aspect of the classical dynamics}
\author{A.~K.~Pogrebkov${}^{*}$\\
${}^{*}$Steklov Mathematical Institute, \\
HSE University, Department of Mathematics,\\
Krichever Center for Advanced Studies at Skoltech;\\ 
Moscowcreation/annihilation of particles.}

\maketitle

\begin{abstract} We show that some quantum effects, such as the creation and annihilation of particles and the existence of dark energy, actually arise in study of known models of classical mechanics, such as the Calogero--Moser and Ruijsenaars-Schneider systems.
\end{abstract}

\section{Introduction}

The development of the theory of integrable systems has led not only to a large number of new, interesting and unexpected results, but also to a revision of a number of established concepts. First of all, we should recall the criticism of the planetary model of the atom. So far, many textbooks of quantum mechanics point to the instability of this model. Indeed, quantum theory provides a stable picture. However, it is also quite strange to talk about the instability of the classical model after the appearance of the theory of solitons. Multi-soliton solutions describe localized wave packets that interact nontrivially with each other, or form bounded states. At the same time, energy and other integrals of motion are preserved and no sinking into a potential hole occurs. Of course, one-dimensional solitons are only a very distant resemblance to the hydrogen atom, but they clearly emphasize the fundamental possibility.

Here we will consider another well-known statement: classical mechanics does not describe the processes of creation and destruction of particles. We show, that in such well-known classical models as the Calogero--Moser (CM) and Ruijsenaars-Schneider (RS) systems, paired particle collisions lead directly to such processes. At the same time, the systems remain integrable, that is, their integrals of motion are always preserved. We also consider some other consequences arising from these properties.

Our consideration here is based on the so-called induced dynamics, a concept introduced in works \cite{AKP_2020_1} and \cite{AKP_2020_2}. This method provides integrable systems by construction, even in cases where there is no Lagrangian and/or Hamiltonian description. Here the dynamical system is given by the zeros of some function. In the Sec.\ 2 we present the main results of this approach. In Sec.\ 3 this method is applied to the description of the dynamics of the singularities of solutions of the KdV and Sinh--Gordon equations. Next, in Sec.\ 4 we consider the CM and RS systems that explicitly demonstrate processes of creation/annihilation of particles. In Sec.\ 5 we consider a system known as "Goldfish" and billiard solution given by this model. Some concluding remarks are given in Sec.\ 6.

\section{Induced dynamics}
In \cite{AKP_2020_1} and \cite{AKP_2020_2} it was suggested to specify dynamical system by means of real zeroes with respect to $x$ of function of $2N$ variables:
\begin{equation}
f(q_1-x,\ldots,q_N-x,p_1,\ldots,p_N)=0,\label{u1}
\end{equation}
where $q_1,\ldots,q_N$ and $p_1,\ldots,p_N$ are canonical variables, 
\begin{equation}
\{q_i,p_j\}=\delta_{ij},\qquad i,j=1,\ldots,N,\label{u2} 
\end{equation}
of a dynamical system of $N$ free particles. We set
\begin{equation}
\dot q_i=h'(p_i),\qquad \dot p_i=0,\quad i=1,\ldots,N.\label{u3}
\end{equation}
Thus we have the Hamiltonian system with respect to the canonical Poisson bracket 
\begin{equation}
q_i=\{q_i,H\},\quad\dot p_i=\{p_i,H\},\quad H=\sum_{i=1}^{N}h(p_i),
\label{u4}
\end{equation}
where $h$ is a function of one variable. We assume that $q_i$ are either real or pairwise complex conjugate and the same are 
properties of the corresponding $p_i$. 

We assume that Eq.\ (\ref{u1}) has $M$ simple real zeros $x_1,\ldots,x_M$, where $0<M\leq N$. Moreover, there exists such open subset of $q_i$'s and $p_j$'s that $M=N$.  

Thus \textbf{induced system} is a system with configuration space given by \textbf{real} zeros 
\begin{equation}
x_i=x_i(q_1(t),\ldots,q_N(t),p_1,\ldots,p_N)\label{u5} 
\end{equation}
of Eq.\ (\ref{u1}). This system is dynamical as all roots $x_{i}$ depend on $t$ via $q_i$ only. Evolution of this system is given by the same Hamiltonian $H$ in (\ref{u4}), 
\begin{equation}
\dot{x_i}=\{x_i,H\},\label{u6}                     
\end{equation}
under the same Poisson bracket (\ref{u2}). This system is integrable by construction and has $N$ independent integrals of motion $p_1.\ldots,p_N$ in involution.

In \cite{AKP_2020_1} we proved existence of the Newton equations for \textbf{the induced dynamical system}, i.e., equations of the type
\begin{equation}
\ddot{x}_{i}=F_{i}(x_1,\ldots,x_N,\dot{x}_1,\ldots,\dot{x}_N),\quad i=1,\ldots,N,\label{u7}
\end{equation}
as well as solvability of the initial value problem. While scheme of solution of the Cauchy problem for the induced system is close to the one for integrable nonlinear PDE's, we do not impose condition of explicit solvability for functions $F_i$.

\section{Dynamics of singularities of the nonlinear integrable equations.}

In works \cite{kdv} and \cite{sinh3}, the dynamics of the singularities of solutions of the KdV, $4u_{t}-6uu_{x}+u_{xxx}=0$, and Sinh--Gordon, $u_{xt}=\frac{1}{16}\sinh u$, equations were studied. In work \cite{AKP_2020_2}, it was shown that the description of the behavior of these singularities, their world lines, is given by induced dynamics. Moreover, this approach is the only way to describe this behavior, since the differential equations of motion for these systems exist only implicitly. In particular, for the Sinh--Gordon equation the world lines of the singularities on the $\{x,t\}$-plane are given as zeroes $x_1(t),\ldots,x_N(t)$ of the product
\begin{equation}
\det(A(x,t)+v)\det(A(x,t)-v)=0, \label{u8}
\end{equation}
where 
\begin{align*}
&A(x,t)=\text{diag}\bigl\{\epsilon_{i}e^{2 p_i(x-q_i)}\bigr\}_{i=1}^{N},\qquad v_{ij}=\dfrac{p_i}{p_{i}+p_{j}},\\
& q_{i}(t)=q_{0,i}-\dfrac{t}{p^{2}_i},\qquad i,j=1,\ldots,N, 
\end{align*}
$A(x,t)$ and $v$ are $N\times N$ matrices, $\Re p_i>0$, and 
\begin{align*}
&\text{either } p_{i}=\overline p_{i}\text{ and }\epsilon_{i}=\pm1, 
\quad q_{i}=\overline q_{i}\\ 
&\text{or }p_{i}=\overline p_{k}\text{ for some }k\neq{i}\text{ and }p_{k}\neq p_{i},\quad \epsilon_{i}=\epsilon_k,\quad q_{i}=\overline q_{k},
\end{align*}
so that Eq.\ (\ref{u8}) is of the kind of (\ref{u1}).
These roots form $N$ smooth time-like curves $x_i(t)$. The first factor gives $u=-\infty$ and the second one 
$u=+\infty$. Lines corresponding to singularities of the different signs can intersect and in these points and only in this 
points they are light-like. This system is Hamiltonian, see (\ref{u6}), where 
\begin{equation}
H=\displaystyle\sum_{i=1}^{N}\dfrac{1}{p_i}.\label{u9}
\end{equation}

Equation of motion for these singularities can be given only in the situation where $N=2$ and in the center of mass, $x_1(t)+x_2(t)=0$:
\begin{equation}
\dfrac{\ddot{x}_{12}\text{sgn}{x_{12}}}{\sqrt{4-{\dot{x}_{12}}^{2}}}=\dfrac{4\varepsilon}{\cosh\biggl(\dfrac{4x_{12}}{\sqrt{4-{
\dot {x}_{12}}^ {2}}}
\sqrt{1+\dfrac{\ddot{x}_{12}\text{}{x_{12}}}{\sqrt{4-{\dot{x}_{12}}^{2}}}}\biggr)-\varepsilon},\label{u10}
\end{equation}
where $x_{12}(t)=x_1(t)-x_2(t)$ and where $\varepsilon=1$ for the case of repulsion and $\varepsilon=-1$ for the both, 
soliton-antisoliton and breather, cases of attraction. We see that $\ddot{x}_{12}$ is involved in both, left and right, hand sides and this dependence is irrational. Nevertheless, this dynamical system is well defined and describes nontrivial interaction of the particles. Description of singulqar solitons in the KdV equation is close to the above, see \cite{AKP_2020_2}. 

\section{Calogero--Moser and Ruijsenaars--Schneider models.}

In the case of CM (see \cite{CM1}--\cite{OP}) and RS (see \cite{RS1}--\cite{RS3}) models differential equations exist and are well known. Say,
\begin{align}
 \text{CM:}\qquad&\ddot{x}_{j}=\sum_{\substack{k=1,\\k\neq j}}^{N}\dfrac{2\gamma^2}{(x_k-x_j)^{3}},\label{u11}\\
 \text{RS:}\qquad&\ddot{x}_{j}=\sum_{\substack{k=1,\\k\neq j}}^{N}\dfrac{2\gamma^2\dot{x}_{j}\dot{x}_{k}}
{(x_j-x_k)\bigl(\gamma^{2}-(x_j-x_k)^{2}\bigr)},\label{u12}
\end{align}
in the rational case. Both systems are completely integrable and have the Lax pairs. In \cite{AKP_2020_2} it was proved that both these models in the rational and hyperbolic cases can be formulated in terms of induced dynamics. Say, the  rational versions of CM and RS models are given by the roots of the equation
\begin{equation}
\det\bigl(Q(t)+W-xI\bigr)=0,\label{u13}
\end{equation}
where $I$ is the unity $N\times N$ matrix,
\begin{equation}
Q(t)=\text{diag}\{q_1(t),\ldots,q_N(t)\},\qquad q_i(t)=a_i+tp_i,\label{u14}                                                       \end{equation}
and
\begin{equation}
W_{\text{CM}}=\biggl(\dfrac{\gamma}{p_{j}-p_{k}}\biggr)_{\substack{j,k=1,\\k\neq j}}^{N},\qquad
W_{\text{RS}}=\biggl(\dfrac{\gamma p_j}{p_{j}-p_{k}}\biggr)_{\substack{j,k=1,\\k\neq j}}^{N}. \label{u15}
\end{equation}

Here again $(q_i,p_j)$ are canonical variables, and Hamiltonian \begin{equation}
H=\sum_{i}p_i^{2}/2\label{u16}                                                                \end{equation}
for \textbf{both these models}. Thus we get another proof of the Liouville integrability for CM and RS systems in the rational case. Moreover, integrability takes place for any choice of the matrix $W(\mathbf{p})$ in (\ref{u13}). 

In contrast to dynamics of singularities for solutions of the KdV and Sinh--Gordon models, CM and RS models have explicit equations of motion. Nevertheless, formulation in terms of induced dynamics enables observation of new properties of these systems.

\subsection{Case $N=2$.}
Thanks to (\ref{u13})--(\ref{u15}) the world lines of the  CM system (\ref{u11}) in this case are given by the equations
\begin{equation}
x_{1,2}(t)=\dfrac{1}{2}\biggl(A+Pt\pm\sqrt{(a+pt)^2-4
\gamma^2p^{-2}}\biggr),\label{u17} 
\end{equation}
where
\begin{equation}
A=a_{1}+a_{2},\qquad a=a_{1}-a_{2},\qquad P=p_{1}+p_{2},\qquad
p=p_{1}-p_{2}.\label{u18}
\end{equation}
We see that in the case where $\gamma^2<0$ system describes two particles under repulsion, see Fig.\ 1. Two particles arrive from negative time infinity, repulse, and fly away to positive time infinity exchanging their momenta.
\begin{figure}[ht]
\begin{multicols}{2}
\hfill
\includegraphics[width=60mm]{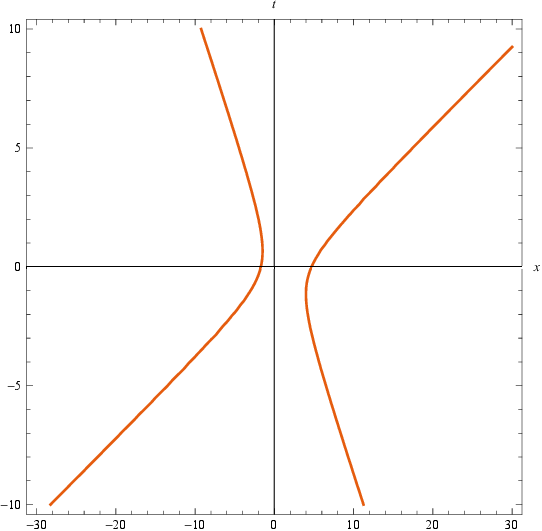}
\hfill
\caption{CM model, 2 particles, $\gamma^2<0$.}
\hfill
\includegraphics[width=60mm]{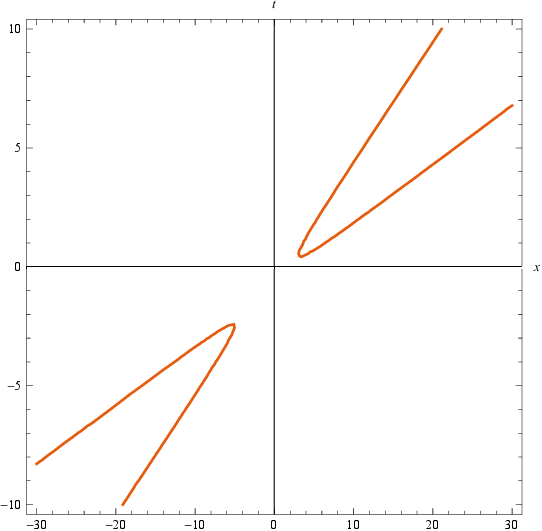}
\hfill
\caption{CM model, 2 particles, $\gamma^2>0$.}
\end{multicols}
\end{figure}
Situation is drastically changed if $\gamma^2>0$. In this case in the interval $\{t_1,t_2\}$,
\begin{equation}
t_1=-ap^{-1}-2\gamma p^{-2},\qquad t_2=-ap^{-1}+2\gamma p^{-2}
\label{u181}
\end{equation}
there exist no any real solution of the system under consideration. Two particles arrive from negative time infinity, attract and annihilate at $t_1$. At moment $t_2$ they appear from nowhere and explodes to infinity, see Fig.\ 2. 

At moments $t_1$ and $t_2$, and at these moments only, particles have infinite velocities. Notice that variables $a_i$ and $p_i$ can be complex, as was mentioned in Sec.\ 1, in contrast to coordinates of particles. Say, in this model variables $a$ and $p$ are pure imaginary if $|(x_1-x_2)(\dot{x}_1-\dot{x}_2)|<2\gamma$, while $x_i(t)$ and $\dot{x}_i(t)$ are real at any moment of $t\notin\{t_1,t_2\}$. We see that times $t_1$ and $t_2$ are real. Moreover, these times are defined by means of the variables $a$ and $p$, so they are movable. This property is similar to the Painlev{\'e} property for differential equations (see, e.g., \cite{MuCo}). The only difference with the case of real $a$ an $p$ is creation at the moment $t_1$ and annihilation at $t_2$, so here particles are unstable and exist only in the interval $\{t_1,t_2\}$. On the other side the initial value problem is solvable in any case where $x_1(0)\neq x_2(0)$ and $|(x_1(0)-x_2(0))(\dot{x}_1(0)-\dot{x}_2(0))|\neq2\gamma$.

\subsection{Multiparticle cases.}
Increasing number of particles preserves their pairwise creation/annihilation. Fig.\ 3 presents behavior of solution of the CM model in the case of 4 particles:
\begin{figure}[ht]
\begin{multicols}{2}
\hfill
\includegraphics[width=60mm]{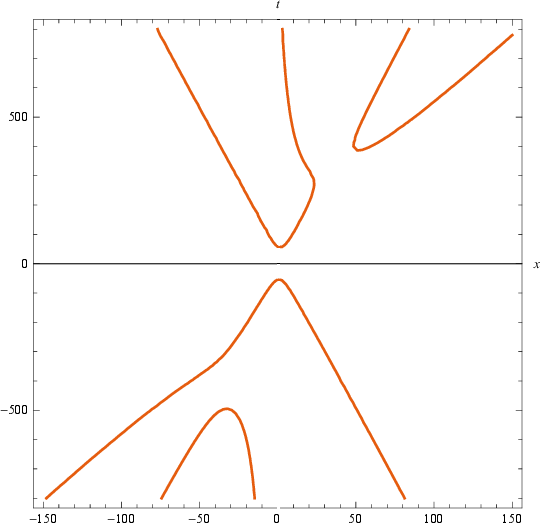}
\hfill
\caption{CM model, 4 particles.}
\hfill
\includegraphics[width=60mm]{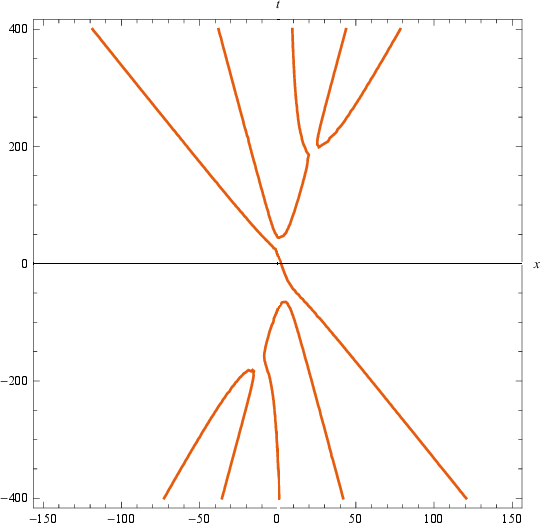}
\hfill
\caption{CM model, 5 particles.}
\end{multicols}
\end{figure}

In the case of odd number of particles at least one particle is preserved, see Fig.\ 4. Thus there exists an interval $\{t_1,t_2\}$ where the system reduces to one particle only. Moreover, we see that the world line of this particle in the interval $\{t_1,t_2\}$ is curved. So it is influenced by those four particles, that ``exist'' in the complex domain only. From the point of view of an outside observer this behavior of the particle in this interval wold look as a curvature of the space.

\section{The ``Goldfish'' model.}

In the previous sections we demonstrated that induced dynamics is useful for the study of dynamical systems. But it is clear that not any dynamical system can be formulated as induced one. One of such systems is the so called 
``Goldfish'' model,
\begin{equation}
\ddot{x}_{j}=2\sum_{\substack{k=1,\\k\neq j}}^{N}\dfrac{\dot{x}_{j}\dot{x}_{k}}{x_j-x_k},\qquad j=1,\ldots,N,\label{u19}
\end{equation}
that was introduced by Calogero in \cite{C1}, \cite{C2}. 

This model can be derived as limit of $\gamma\to\infty$ of the rational RS model (\ref{u12}). It has Lax pair with $L$-operator
\begin{equation}
L=(1,\ldots,1)^{\text{T}}\otimes(\dot{x}_1,\ldots,\dot{x}_N),
\label{u20}
\end{equation}
that is degenerate, $L^{n}=\bigl(\sum_{j=1}^{N}\dot{x}_j\bigr)^{n-1}L$, so it do not generate integrals of motion. Nevertheless, this system is integrable, it has proper number of integrals of motion in involution and obeys extremely rich symmetry: $x\to\alpha+\lambda x$, $t\to\beta+\mu t$, where $\alpha,\beta,\lambda,\mu$ are arbitrary constants. In particular this model is Lorentz invariant. Besides this, any particle initially at rest maintains this state of rest forever---since $\dot{x}_n=0$ implies $\ddot{x}_n=0$ and so on. 

``Goldfish'' model has no formulation in terms of induced dynamics, but its solution can be defined by means of the algebraic equation in terms of the initial data. Calogero proved that values $x_n(t)$ of the $N$ particles in this system obey the following equation in $x$:
\begin{equation}
\sum_{j=1}^{N}\dfrac{\dot{x}_j(0)}{x-x_j(0)}=\dfrac{1}{t}.
\end{equation}
In the case where initial velocities $\dot{x}_1(0)$ and $\dot{x}_2(0)$ have opposite signs this system has solution as on Fig.\ 2 with creation/annihilation of particles. On the other side Lorenz-invariance of this system impose condition that $\dot{x}_n(t)>0$ for any $n$ and at any $t$. So it is reasonable to consider solution with nonegative initial velocities. In the Fig.\ 5 the world lines for the case of six particles is demonstrated. We see that five of these particles are moving at almost zero velocities, and the six particle appeared on the left. It scatters on the first of the five particles and flies away at velocity close to zero. The first of the five particles scatters on the second one, and so on, up to the last particle, which flies away at infinity.
\begin{figure}[ht]
\noindent\centering{
\includegraphics[width=60mm]{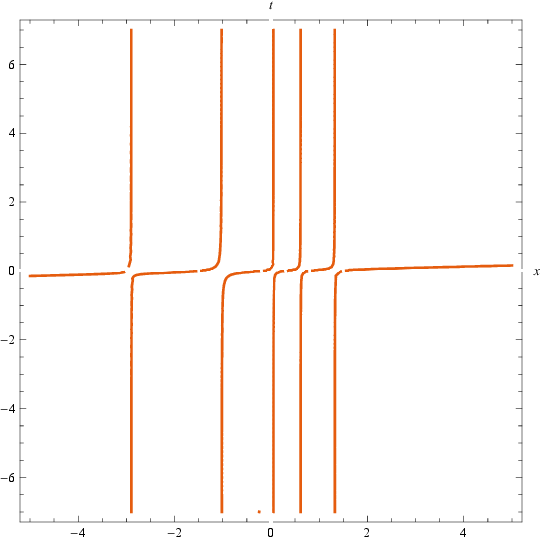}}
\caption{``Goldfish'' model, 6 particles}
\end{figure}
Thus, we have an example of a system of particles behaving like a rigid balls in billiard. 

\section{Conclusion.}
It was shown above that in the case of integrable system classical dynamics can describe processes of \textbf{creation and annihilation} of particles. At the same time, the energy of the system remains unchanged under these processes, as well as values of all integrals of motion. The particles are transformed from the observed state to the state of \textbf{dark matter} and vice versa. Fig.\ 3 shows that the annihilated particles do not fall out of the system. The remaining particles feel their presence.  It is clear that we can create a system of a huge number of particles, where all of them disappear almost simultaneously, and then appear almost simultaneously. From the point of view of an outside observer, such a phenomenon would look like a \textbf{big bang}. Moreover, Figs.\ 2, 3, and 4 show that the particles are flying apart relative to each other---the effect of an \textbf{expanding universe}.

Some analogy between quantum theory and the theory of classical integrable systems is not surprising. Both theories are based on the spectral properties of differential operators and often use the same methods and results. We are not saying here that classical theory can replace the quantum one. Our task is to draw attention to the effects described above. In particular, it would be useful to consider three-dimensional analogues of these effects, as well as quantization on the manifold of such classical solutions.

\end{document}